\documentclass{article}
\usepackage{ijcai20}

\usepackage{times}
\usepackage{soul}
\usepackage{url}
\usepackage[hidelinks]{hyperref}
\usepackage[utf8]{inputenc}
\usepackage[small]{caption}
\usepackage{graphicx}
\usepackage{amsmath}
\usepackage{amsthm}
\usepackage{booktabs}
\usepackage{algorithm}
\usepackage{algorithmic}
\usepackage{amsfonts}
\usepackage{booktabs}
\usepackage{multirow}
\usepackage{algorithm, algorithmic}
\usepackage{amssymb}
\usepackage{xspace}

\newcommand{\ie}{\emph{i.e.,}\xspace}
\newcommand{\eg}{\emph{e.g.,}\xspace}

\def\B#1{\boldsymbol #1}
\def\C#1{\mathcal #1}

\title{Contextualized Graph Attention Network for Recommendation with Item Knowledge Graph}

\author{
 Susen Yang$^1$\and
 Yong Liu$^2$\and
 Yonghui Xu$^2$\and
 Chunyan Miao$^2$\and
 Min Wu$^3$\and
 Juyong Zhang$^1$
 \affiliations
 $^1$University of Science and Technology of China, China\\
 $^2$Nanyang Technological University, Singapore\\
 $^3$Institute for Infocomm Research, A$^\ast$STAR, Singapore\\
 \emails
 $^1$\{susen@mail., juyong@\}ustc.edu.cn, $^2$\{stephenliu, xuyh, ascymiao\}@ntu.edu.sg, $^3$wumin@i2r.a-star.edu.sg
}

\begin{document}

\maketitle

\begin{abstract}
Graph neural networks (GNN) have recently been applied to exploit knowledge graph (KG) for recommendation. Existing GNN-based methods explicitly model the dependency between an entity and its local graph context in KG (\ie the set of its first-order neighbors), but may not be effective in capturing its non-local graph context (\ie the set of most related high-order neighbors). In this paper, we propose a novel recommendation framework, named Contextualized Graph Attention Network (CGAT), which can explicitly exploit both local and non-local graph context information of an entity in KG. Specifically, CGAT captures the local context information by a user-specific graph attention mechanism, considering a user's personalized preferences on entities. Moreover, CGAT employs a biased random walk sampling process to extract the non-local context of an entity, and utilizes a Recurrent Neural Network (RNN) to model the dependency between the entity and its non-local contextual entities. To capture the user's personalized preferences on items, an item-specific attention mechanism is also developed to model the dependency between a target item and the contextual items extracted from the user's historical behaviors. Experimental results on real datasets demonstrate the effectiveness of CGAT, compared with state-of-the-art KG-based recommendation methods.
\end{abstract}

\section{Introduction}
Personalized recommender systems have been widely applied in different application scenarios~\cite{liu2014exploiting,liu2017learning,liu2018dynamic,wu2019pd}. The knowledge graph (KG) including rich semantic relations between items has recently been shown to be effective in improving recommendation performances~\cite{sun2019research}. Essentially, KG is a heterogeneous network where nodes correspond to entities and edges correspond to relations. The main challenge of incorporating KG for recommendation is how to effectively exploit the relations between entities and the graph structure of KG. In practice, one group of methods impose well-designed additive regularization loss term to capture the KG structure~\cite{zhang2016collaborative,cao2019unifying}. However, they can not explicitly consider the semantic relation information of KG into the recommendation model. Another group of methods focus on extracting the high-order connectivity information between entities along paths which are always manually designed or selected based on special criteria~\cite{yu2013recommendation,zhao2017meta}. These approaches may heavily rely on domain knowledge. Recently, the quick development of graph neural networks (GNN)~\cite{zhou2018graph} motivates the application of graph convolutional networks (GCN)~\cite{kipf2016semi} and graph attention networks (GAT)~\cite{velivckovic2017graph} in developing end-to-end KG-based recommender systems~\cite{wang2019knowledge,wang2019kgat}, which can aggregate the context information from the structural neighbors of an entity in KG.

\begin{figure}
\centering
\includegraphics[width=0.85\columnwidth]{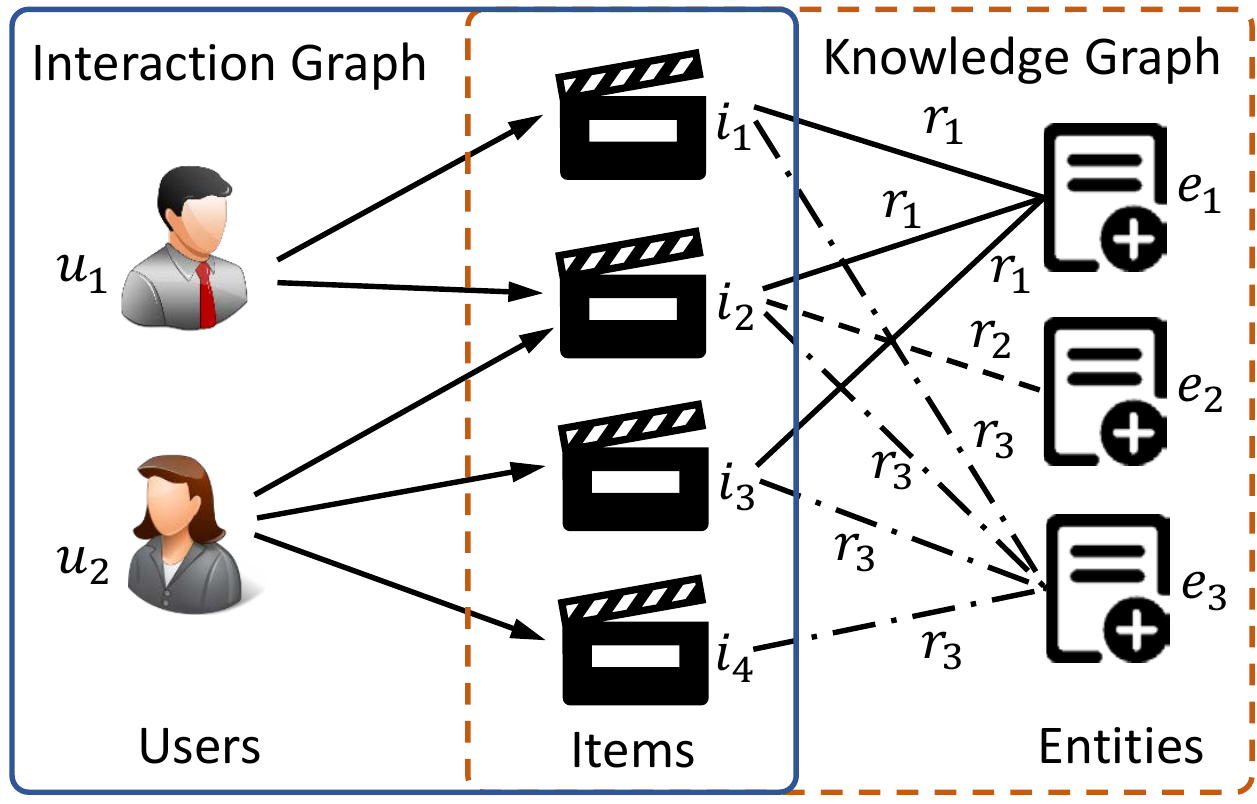}
%\vspace{-5pt}
\caption{A simple example showing the user-item interactions and the item knowledge graph.}
%\vspace{-15pt}
\label{fig:fig1}
\end{figure}

Although GNN-based recommendation methods can automatically capture both the structure and semantic information of KG, they may still have the following deficiencies. Firstly, most GNN-based methods lack of modeling user-specific preferences on entities, when aggregating the \textbf{\emph{local graph context}} (\ie the first-order neighbors) of an entity in KG. As shown in Figure~\ref{fig:fig1}, both users have interactions with the item $i_{2}$. However, they prefer $i_2$ may due to different reasons. For example, $u_1$ prefers $i_2$ because of the attribute entity $e_1$ of $i_2$ in KG, while $u_2$ pays more attentions to its attribute entity $e_3$. The methods that ignore this situation are insufficient to model users' personalized preferences. Secondly, the \textbf{\emph{non-local graph context}} (\ie the set of most related high-order neighbors) of an entity in KG is not explicitly captured in existing GNN-based recommendation methods. In KG, some items may have very few neighbors, thus some important entities may not be directly connected to them. For example, in Figure~\ref{fig:fig1}, the item $i_4$ has only one entity $e_3$ linked with it, thus the aggregation of local context information for the entity $e_3$ is not enough to represent $i_4$. Moreover, we can also observe that entity $e_1$ is connected with $i_4$ along many multi-hop paths, which demonstrates the importance of $e_1$ to $i_4$. Exiting GNN-based methods~\cite{wang2019knowledge,wang2019kgat} address this limitation by feature propagation layer by layer. However, this may weaken the effects of farther connected entities or even bring noise information.

To address these issues, we propose a novel recommendation framework, namely Contextualized Graph Attention Network (CGAT), which explicitly exploits both the local and non-local context of an entity in KG, as well as the item context extracted from users' historical data. The contributions made in this paper are as follows: (1) We propose a user-specific graph attention mechanism to aggregate the local context information in KG for recommendation, based on the intuition that different users may have different preferences on the same entity in KG; (2) We propose to explicitly exploit the non-local context information in KG, by developing a biased random walk sampling process to extract the non-local context of an entity, and employing a recurrent neural network (RNN) to model the dependency between the entity and its non-local context in KG; (3) We develop an item-specific attention mechanism that exploits the context information extracted from a user's historical behavior data to model her preferences on items; (4) We perform extensive experiments on real datasets to demonstrate the effectiveness of CGAT. Experimental results indicate that CGAT usually outperforms state-of-the-art KG-based recommendation methods.

\section{Related Work}
KG-based recommendation methods can be categorized into three main groups: regularization-based methods, path-based methods, and GNN-based methods. The regularization-based methods exploit the KG structure by imposing regularization terms into the loss function used to learn entity embedding. For example, CKE~\cite{zhang2016collaborative} is a representative method, which uses TransR~\cite{lin2015learning} to derive semantic entity representations from item KG. %These representations are then used to enhance recommendation performances.
The KTUP model~\cite{cao2019unifying} is proposed to jointly train the personalized recommendation and KG completion tasks, by sharing the item embedding. The high-order feature interactions between items and entities can be further approximated by a cross\&compress unit~\cite{wang2019multi}. These methods are highly flexible. However, they lack an explicit modeling of the semantic relations in KG. The path-based methods exploit various connection patterns between entities. For example, the recent works~\cite{yu2013recommendation,shi2015semantic} estimate the meta-path based similarities for recommendation. In~\cite{zhao2017meta}, matrix factorization and factorization machine techniques are integrated to assemble different meta-path information. To address the limitation of manually designed meta-paths, different selection rules or propagation methods have been proposed~\cite{wang2018ripplenet}. For example, in~\cite{sun2018recurrent}, the length condition is used to extract paths and then a batch of RNN are applied to aggregate the path information. Besides the length, multi-hop relational paths can also be inducted based on item associations~\cite{ma2019jointly}. Recently, the GNN-based methods aim to develop the end-to-end KG-based recommender systems.
%For example, the GC-MC model~\cite{berg2017graph} considers the user-item interaction data as a bipartite graph, and applies a GCN encoder to this graph to learn the user and item features. A bilinear decoder is then employed to predict missing ratings.
%The DKN model~\cite{wang2018dkn} associates each word in the news content with relevant entities, and incorporates the KG representation into news recommendation. A knowledge-aware CNN model is proposed to fuse the words and entities embedding by considering them as multiple stacked channels.
For example, the KGNN-LS model~\cite{wang2019knowledge} employs a trainable function that calculates the relation weights for each user to transfer the KG into a user-specific weighted graph, and then applies GCN on this graph to learn item embedding. In~\cite{wang2019kgat}, the graph attention mechanism is adopted to aggregate and propagate local neighborhood information of an entity, without considering users' personalized preferences on entities. On summary, these GNN-based methods implicitly aggregate the high-order neighborhood information via layer by layer propagation, instead of explicitly modeling the dependency between an entity and its high-order neighbors.

\begin{figure*}[t]
\centering
\includegraphics[width=0.9\textwidth]{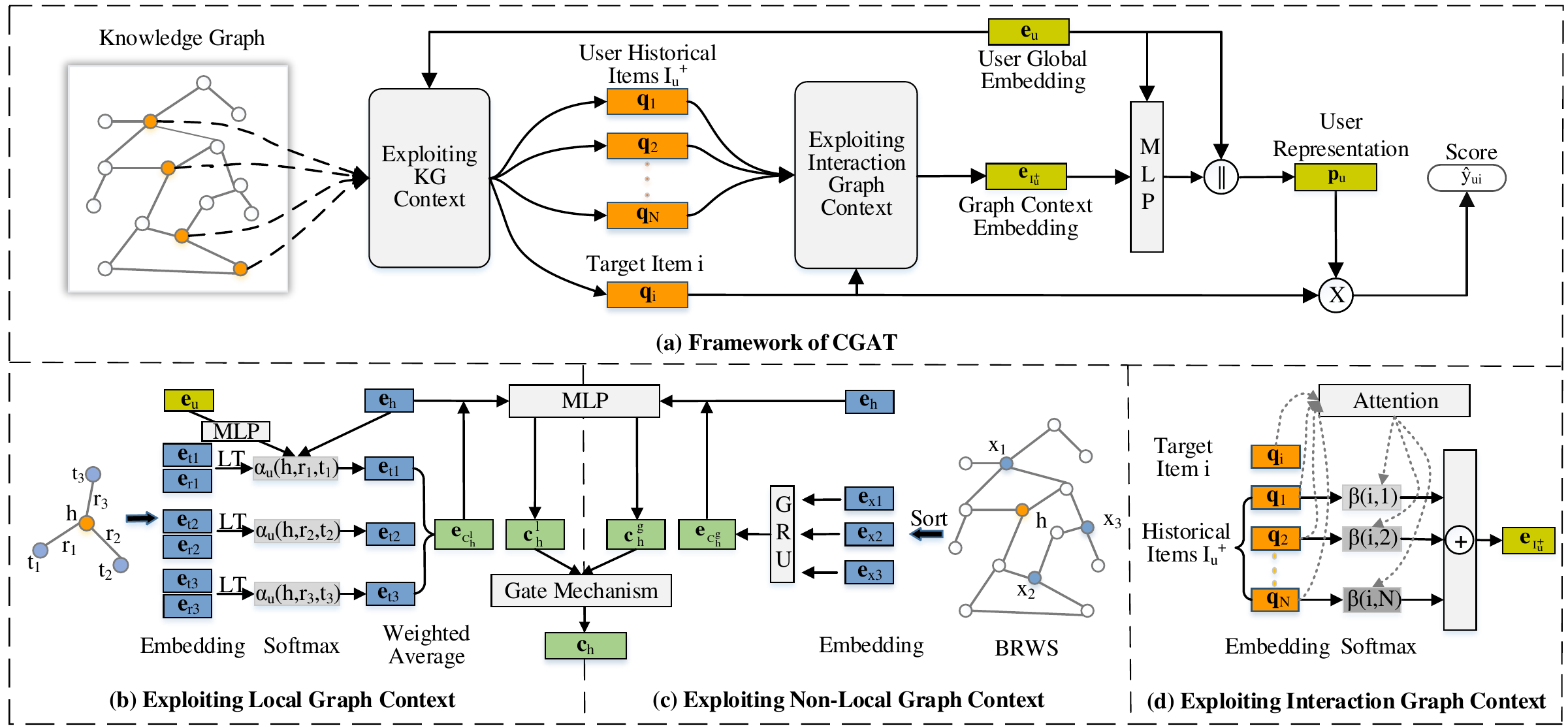} % Reduce the figure size so that it is slightly narrower than the column.
%\vspace{-8pt}
\caption{(a) The framework of the CGAT. From left to right, it exploits the KG context and interaction graph context to predict a user's preference score on a candidate item; (b) Exploiting local graph context by applying a user-specific graph attention mechanism to KG; (c) Exploiting non-local graph context by bias random walk based sampling (BRWS) and GRU module; (d) Exploiting interaction graph context by applying an item-specific attention mechanism to the user's historical items.}
\vspace{-13pt}
\label{fig:mymodel}
\end{figure*}

\section{Contextualized Graph Attention Network}
We assume the item KG $\C{G}=\{\C{E}, \C{R}, \C{D}\}$ is available, where $\C{E}$ denotes the set of entities, $\C{R}$ denotes the set of relations, and $\C{D}$ denotes the set of entity-relation-entity triples $(h, r, t)$ describing the KG structure. Here $h\in\mathcal{E}$, $r\in\mathcal{R}$, and $t\in\mathcal{E}$ denote the head entity, relation, and tail entity of a knowledge triple, respectively. $\B{e}_h \in \mathbb{R}^{1 \times d}$ and $\B{e}_r \in \mathbb{R}^{1 \times d}$ are used to denote the embedding of the entity $h$ and relation $r$, where $d$ denotes the dimensionality of latent space. Note that the items are treated as a special type of entities in the KG. In addition, we denote the set of users by $\C{U}$, the set of items by $\C{I}$, and all the observed user-item interactions by $\C{O}$. For each user $u$, we denote the set of items she has interacted by $\C{I}_{u}^{+}$, and use $\B{e}_u \in \mathbb{R}^{1\times d}$ to denote her embedding.
%This paper proposes the CGAT model to effectively exploit the KG $\C{G}$ and users' historical interaction data $\C{O}$ for recommendation.
Figure~\ref{fig:mymodel} shows the structure details of the proposed CGAT model.

\subsection{Exploiting Knowledge Graph Context}
CGAT exploits KG context from two aspects: (a) local context information, and (b) non-local context information.

\subsubsection{Local Graph Context}
For the entity corresponding to an item, it is always linked with many other entities that can enrich its information in KG.
%Different methods, e.g., GCN~\cite{kipf2016semi} and GAT~\cite{velivckovic2017graph}, can be used to aggregate the neighborhood information of an entity. However, these methods usually ignore the user's personalized preferences on entities.
To consider users' personalized preferences on entities, we develop a user-specific graph attention mechanism to aggregate the neighborhood information of an entity in KG. For different users, we compute different attention scores for the same neighborhood entity. The embedding of neighborhood entities can then be aggregated based on the user-specific attention scores. Here, we denote the local neighbors of an entity $h$ by $\C{C}_{h}^l = \{t|(h, r, t)\in\C{D}\}$, and define $\C{C}_{h}^l$ as the \textbf{\emph{local graph context}} of $h$ in KG. Moreover, we also argue that the neighborhood entities may have different impacts, if they are connected via different relations.
%For example, the director and actor of a movie may be the same person. This person usually have different impacts on the movie from the actor perspective and the director perspective.
To incorporate relation into the attention mechanism, we firstly integrate the embedding of a neighborhood entity $t\in \mathcal{C}_h^l$ and the embedding of corresponding relation $r$ by the following linear transformation,
\begin{equation}
\B{e}_{rt} = \big(\B{e}_{r}||\B{e}_{t}\big)\B{W}_{0},
\label{eq:relationagg}
\end{equation}
where $||$ is the concatenation operation, $\B{W}_{0} \in \mathbb{R}^{2d \times d}$ is the weight matrix. The user-specific attention score $\alpha_{u}(h,r,t)$ that describes the importance of the entity $t \in \mathcal{C}_h^{l}$ to the entity $h$, for a target user $u$, is defined as follows,
\begin{equation}
\alpha_u(h,r,t) = \frac{\exp\big[\pi_u(h,r,t)\big]}{\sum_{(h,\tilde{r},\tilde{t})\in\C{D}}\exp\big[\pi_u(h,\tilde{r},\tilde{t})\big]}.
\label{eq:localattention}
\end{equation}
The operation $\pi_u(h,r,t)$ is performed by a single-layer feed forward neural network, which is defined as follows,
%\begin{equation}
%\pi_u(h,r,t) = \mbox{tanh}\big[(\B{e}_h || \B{e}_{rt})\B{W}_{1}+ \B{b}_1 \big]\B{e}_u^{\top},
%\end{equation}
\begin{equation}
\pi_u(h,r,t) = \mbox{tanh}\big[(\B{e}_h || \B{e}_{rt})\B{W}_{1}+ \B{b}_1 \big]\B{m}_u^{\top},
\label{eq:attention3}
\end{equation}
where $\B{m}_u$ is a non-linear transform of $\B{e}_u$ defined as $\B{m}_u=\mbox{ReLU}(\B{e}_u\widetilde{\B{W}}_1 + \widetilde{\B{b}}_1)$. %where we choose tanh~\cite{velivckovic2017graph} as the nonlinear activation function,
Here, $\widetilde{\B{W}}_1 \in \mathbb{R}^{d\times d}$, $\B{W}_1 \in \mathbb{R}^{2d\times d}$, and $\B{b}_1, \widetilde{\B{b}}_1 \in \mathbb{R}^{1\times d}$ are the weight matrices and bias vectors respectively. %The usage of $\B{e}_u$ makes the attention mechanism become user-specific.
Given the coefficient of each neighboring entity of $h$, we compute the linear combination of their embedding to obtain the local neighborhood embedding of $h$ as follows,
\begin{equation}
\B{e}_{\C{C}_h^{l}}=\sum_{t\in \C{C}_h^l}\alpha_u(h,r,t) \B{e}_{t}.
\end{equation}
Then, we aggregate the embedding of entity $h$ and it's local neighborhood embedding $\B{e}_{\C{C}_h^{l}}$ to form a local contextual embedding $\B{c}^{l}_{h}$ for $h$ as follows,
\begin{equation}
\B{c}^{l}_h = \mbox{tanh}\big[(\B{e}_h || \B{e}_{\C{C}_h^{l}})\B{W}_{2}+ \B{b}_{2}\big],
\label{eq:localagg}
\end{equation}
where $\B{W}_{2} \in \mathbb{R}^{2d\times d}$ and $\B{b}_{2} \in \mathbb{R}^{1\times d}$ are the weight matrix and bias vector of the
aggregator.

\subsubsection{Non-Local Graph Context}
The user-specific graph attention network explicitly aggregates the local neighbor (one-hop) information of a target entity to enrich the representation of the target entity. However, this is not enough to capture the non-local context of an entity in KG, and also has weak representation ability for the nodes which have few connections in KG. To offset this gap, we propose a biased random walk based GRU module to aggregate non-local context information of entities.

The biased random walk sampling (BRWS) procedure is used to extract the non-local context of a target entity $h$. To achieve a wider depth-first search, we repeat biased random walk from $h$ to obtain $M$ paths, which have a fixed length $L$. The walk iteratively travels to the neighbors of current entity with a probability $p$, which is defined as follows,
\begin{equation}
    p(t_{k+1}) =
    \begin{cases}
    \gamma & \mbox{if~} t_{k+1} \in \C{C}^l_{t_{k-1}} \mbox{or}~ t_{k+1}=t_{k-1},\\
    1-\gamma & \mbox{else},
    \end{cases}
\end{equation}
where $t_k$ is the $k$-th entity of a path, $t_0$ denotes the root entity $h$. To encourage wider search, we empirically set $0 < \gamma < 0.5$. After obtaining the $M$ paths and $M*L$ entities by walk, we sort entities according to their frequency in walks in descending order, and choose a set of top-ranked entities orderly. These entities are defined as the \textbf{\textit{non-local graph context}} of the entity $h$ in KG, and denoted by $\C{C}^g_{h}$. In the experiments, we empirically set $|\C{C}_{h}^{g}|=|\C{C}_{h}^{l}|$, and set the parameters $\gamma$, $M$, and $L$ to 0.2, 15, and 8, respectively.

%GRU is a variant of LSTM model, which has been successfully used for various sequential recommendation tasks.
In this work, we employ GRU to model the dependency between an entity $h$ and its non-local context $\C{C}_{h}^{g}$, because GRU can yield better performance in processing sequence data (\ie $\mathcal{C}^g_{h}$ can be seen as a frequency sequence data). Indeed, the more frequently an entity appears in random walks, the more important it is to the target entity $h$. Based on this intuition, we input $\mathcal{C}^g_{h}$ into GRU in reverse order, and use the last step output as the embedding of $\mathcal{C}^g_{h}$, which is denoted by,
\begin{equation}
\B{e}_{\C{C}^g_{h}} = \mbox{GRU}\big(\overleftarrow{\C{C}^g_h}\big),
\end{equation}
where $\overleftarrow{\C{C}^g_h}$ denotes the reverse set of $\C{C}^g_h$. Then, we aggregate $\B{e}_{h}$ and $\B{e}_{\mathcal{C}^g_{h}}$ to form the non-local contextual embedding $\B{c}_{h}^{g}$ for $h$ as follows,
\begin{equation}
\B{c}^{g}_h = \mbox{tanh}\big[(\B{e}_h || \B{e}_{\C{C}_h^{g}})\B{W}_{2} + \B{b}_{2}\big].
\label{eq:globalagg}
\end{equation}
Here, we use the same aggregator parameters as in Eq~\eqref{eq:localagg}. Given the embeddings of local and non-local context of $h$ in KG, we apply a gate mechanism to integrate these two embeddings by learning the weights in each dimension as,
\begin{equation}
\B{c}_{h} = \sigma(\B{\omega}) \odot \B{c}^{l}_h + (1-\sigma(\B{\omega})) \odot \B{c}^{g}_h,
\label{eq:kgcontextemb}
\end{equation}
where $\B{\omega} \in \mathbb{R}^{1\times d}$ is a learnable vector, $\sigma(\cdot)$ denotes the sigmoid function. As items are a special type of entities in KG, we can use Eq.~\eqref{eq:kgcontextemb} to compute the context embedding $\B{c}_i$ of item $i$, considering its local context $\C{C}_{i}^{l}$ and non-local context $\C{C}_{i}^{g}$ in KG. Then, we concatenate $\B{e}_i$ and $\B{c}_i$ to obtain the contextualized representation of an item $i$ as $\B{q}_i = (\B{e}_i||\B{c}_i)$.

%\begin{equation}
%\B{q}_i = (\B{e}_i||\B{c}_i).
%\label{eq:itemembeddings}
%\end{equation}

\subsection{Exploiting Interaction Graph Context}
In practice, a user's historical items are usually used to describe her potential interests~\cite{shi2014collaborative}. For example, the classical SVD++ model~\cite{koren2008factorization} treats a user $u$'s historical items $\C{I}_{u}^{+}$ as the implicit feedback given by $u$, and model the influences of $\C{I}_{u}^{+}$ on a target item $i$ for recommendation. Following similar spirit, we define $\C{I}_{u}^{+}$ as the \textbf{\emph{interaction graph context}} of user $u$. Then, we develop an item-specific attention mechanism to model the influences of $\C{I}_{u}^{+}$ on $i$. The basic assumption is that a user's historical item may have different importance in estimating her preferences on different candidate items. For each item $j \in \C{I}_u^+$, its relevance weight with respect to the target item $i$ is defined as,
\begin{align}
\beta(i, j) = \frac{\exp\big[ \mbox{tanh}\big((\B{q}_i || \B{q}_{j})\B{w}^{\top}+b\big) \big]}{\sum_{k\in \C{I}_u^+}\exp\big[ \mbox{tanh}\big((\B{q}_i || \B{q}_{k})\B{w}^{\top}+b\big) \big]},
\label{eq:interactionattention}
\end{align}
where $\B{w} \in \mathbb{R}^{1 \times 4d}$ is a weight vector, $b$ is the bias, $\B{q}_i$ and $\B{q}_{j}$ are the contextualized representations of items $i$ and $j$. Then, we define the embedding of the graph context $\C{I}_{u}^{+}$, with respect to a target item $i$, as follows,
\begin{equation}
\B{e}_{\C{I}_{u}^+} = \sum_{j\in\C{I}_{u}^+}\beta(i, j)\B{q}_j.
\end{equation}
%Note that this context embedding is \emph{item-specific}, it can be seen as $u$'s local interest representation for the target item $i$. In other words, for different candidate item, $\B{c}_{\C{I}_{u}^+}$ may be different.
A non-linear transformation, where ReLU is the activation function, is then used to aggregate $\B{e}_u$ and $\B{e}_{\C{I}_{u}^+}$ to form the contextual embedding for $u$ as follows,
\begin{equation}
\B{c}_u = \mbox{ReLU}\big[(\B{e}_u||\B{e}_{\C{I}_{u}^+})\B{W}_3 + \B{b}_3\big],
\label{eq:userembeddings}
\end{equation}
where $\B{W}_3\in \mathbb{R}^{3d \times d}$ and $\B{b}_3\in \mathbb{R}^{1 \times d}$ are the weight matrix and bias vector. We concatenate $\B{e}_u$ and $\B{c}_u$ to form the contextualized representation for $u$ as $\B{p}_u = (\B{e}_u||\B{c}_u)$. The prediction of $u$'s preference on $i$ can be defined as $\hat{y}_{ui} = {\B{p}_u}\B{q}_i^\top$.

\begin{algorithm}[t]
    \label{alg:optimization}
    \renewcommand{\algorithmicrequire}{\textbf{Input:}}
    \renewcommand{\algorithmicensure}{\textbf{Output:}}
    \caption{CGAT Optimization Algorithm}
    \label{alg:1}
    \begin{algorithmic}[1]
        \REQUIRE Observed interactions $\C{O}$, knowledge graph $\C{G}$
        \ENSURE Score function $\C{F}(u,i;\B{\Theta})=\hat{y}_{ui}$
        \STATE Randomly initialize all parameters
        \STATE Construct the set $\widetilde{\C{O}}$ and $\widetilde{\mathcal{D}}$ based on $\C{O}$ and $\C{D}$;
        \FOR{$iter = 1, 2, \cdots, max\_iter$}
        \STATE Sample a batch of tuples $\C{B}_{1}$ from $\widetilde{\C{O}}$;
        \STATE Sample a batch of tuples $\C{B}_{2}$ from $\widetilde{\mathcal{D}}$;
        %\STATE Exploit interaction context for users in $\C{B}_{1}$ and KG context for items in $\C{B}_{2}$;
        \STATE Compute gradients of Eq.~\eqref{eq:finalloss} with respect to $\B{\Theta}$ by back-propagation, based on tuples in $\C{B}_1$ and $\C{B}_{2}$;
        \STATE Update $\B{\Theta}$ by gradient descent algorithm (\ie Adam) with learning rate $\eta$;
        \ENDFOR
        \RETURN{$\C{F}(u,i;\B{\Theta})$}
    \end{algorithmic}
\end{algorithm}

\subsection{Learning Algorithm}
The Bayesian personalized ranking (BPR) optimization criterion~\cite{rendle2009bpr} is used to learn the model parameters of CGAT. BPR assumes that the interacted items should have higher ranking scores than the un-interacted items for each user. Here, we define the BPR loss as follows,
\begin{equation}
\C{L}_{\mbox{BPR}}=\sum_{(u,i^{+},i^{-})\in\widetilde{\C{O}}}-\log\sigma(\hat{y}_{ui^+}-\hat{y}_{ui^-}),
\end{equation}
where $\widetilde{\C{O}}$ is constructed by negative sampling. Empirically, for each $(u, i)\in \C{O}$, we randomly sampling $5$ items from $\C{I} \setminus \C{I}_u^{+}$ in the experiments. As we also need to learn the embedding of entities and relations in KG, we design a regularization loss based on the KG structure. %to make the proposed CGAT model avoid over-fitting.
Specifically, for each triple $(h,r,t) \in \C{D}$, we first define the following score to describe the distance between the head entity $h$ and the tail entity $t$ via relation $r$ in the latent space,
\begin{equation}
s_r(h,t)=||\B{e}_h-\B{e}_{rt}||^2_2.
\end{equation}
Then, we define the regularization loss as follows,
\begin{equation}
\C{L}_{\mbox{KG}}=\sum_{(h,r,t,t') \in \widetilde{\mathcal{D}}} \log\sigma\big(s_r(h,t)-s_r(h,t')\big),
\label{eq:kgloss}
\end{equation}
where $\widetilde{\mathcal{D}}$ is constructed by randomly sampling an entity $t'$ from $\C{E}\setminus \C{C}_{h}^{l}$, for each $(h,r,t)\in\C{D}$. The motivation is that, in the latent space, the distance between an entity $h$ and its directly connected neighbor $t$ should be smaller than the distance between $h$ and the entity $t'$ that is not directly connected to $h$, via relation $r$. Then, the model parameters can be learned by solving the following objective function,
\begin{align}
\min_{\B{\Theta}} \C{L}_{\mbox{BPR}} + \lambda_{1} \mathcal{L}_{\mbox{KG}} + \lambda_2 ||\B{\Theta}||^2_2,
\label{eq:finalloss}
\end{align}
where $\B{\Theta}$ denotes all the parameters of CGAT,  $\lambda_1$ and $\lambda_2$ are the regularization parameters. The problem in Eq.~\eqref{eq:finalloss} is solved by a gradient descent algorithm. The details of the optimization algorithm are summarized in Algorithm 1.

In the implementation of CGAT, we randomly sample $S$ neighbors from $\C{C}_{h}^{l}$ for a target entity $h$, and $N$ historical items from $\C{I_{u}^+}$ for a target user $u$, to compute the attention weights defined in Eq.~\eqref{eq:localattention} and Eq.~\eqref{eq:interactionattention} respectively. This trick can help keep the computational pattern of each mini-batch fixed and improve the computation efficiency. Moreover, we also set the size of non-local context $|\C{C}_{h}^{g}|$ to $S$. In model training, $S$ and $N$ are fixed. Let $B$ denote the number of sampled user-item interactions in each batch. The time complexity of biased random walk sampling procedure is $O(|\C{I}|SML)$, which can be performed before training. In each iteration, to exploit KG context, the user-specific graph attention mechanism and the GRU module have computational complexity $O(BNSd^2)$. The complexity of exploiting interaction graph context is $O(BNd^2)$. The overall complexity of each mini-bacth iteration is $O\big(B(NSd^2+Nd^2)\big) \approx O(BNSd^2)$, which is linear with all hyper-parameters except for $d$. %Thus, the proposed CGAT method is very efficient. %than traditional GCN-based methods that have the exponentially increasing time complexity with the number of propagation layers.

\section{Experiments}

\subsection{Experimental Settings}
\noindent\textbf{Datasets}: The experiments are performed on three public datasets: Last-FM\footnote{https://grouplens.org/datasets/hetrec-2011/}, Movielens-1M\footnote{https://grouplens.org/datasets/movielens/1m/}, and Book-Crossing\footnote{http://www2.informatik.uni-freiburg.de/$\sim$cziegler/BX/} (respectively denoted by FM, ML, and BC). Following~\cite{wang2018ripplenet,wang2019multi,wang2019knowledge}, we keep all the ratings on FM and BC datasets as observed implicit feedback, due to data sparsity. For ML dataset, we keep ratings larger than 4 as implicit feedback. The KGs of these datasets are constructed by Microsoft Satori, and are currently public available\footnote{https://github.com/hwwang55}. As introduced in~\cite{wang2019multi}, only the triples from the whole KG with a confidence level greater than 0.9 are retained. The sizes of ML and BC KGs are further reduced by only selecting the triples where the relation name contains "film" and "book", respectively. For these datasets, we match the items and entities in sub-KGs by their names (\eg head, film.film.name, tail for ML). The items matching no entities or multiple entities are removed. Table~\ref{tab:tab2} summarizes the statistics of these experimental datasets.
\begin{table}
    \centering
    \small
    \caption{Statistics of the experimental datasets.}
    %\vspace{-8pt}
    \label{tab:tab2}
    \begin{tabular}{l|c|c|c} \hline
    & FM & ML & BC \\ \hline
    \#Users & 1,872 & 6,036 & 17,860 \\%& 63,566\\
    \#Items & 3,846  & 2,347  & 14,967 \\%& 1,362  \\
    \#Interactions & 21,173 & 376,886 & 69,876 \\%& 3,257,131 \\
    \#Density & 0.29\% & 2.66\% & 0.026\% \\\hline%& 3.76\%\\ \hline
    \#Entities & 9,366 & 7,008 &  77,903 \\%& 28,115\\
    \#Relations & 60 & 7 & 25 \\%& 7\\
    \#Triples & 15,518 & 20,782 & 151,500 \\%& 160,519 \\
    \hline
    \end{tabular}
    %\vspace{-10pt}
\end{table}

\noindent\textbf{Setup and Metrics}: For each dataset, we randomly select 60\% of the observed user-item interactions for model training, and choose another 20\% of interactions for parameter tuning. The remaining 20\% of interactions are used as testing data. %to evaluate the effectiveness of recommendation methods.
The quality of the top-$K$ item recommendation is assessed by three widely used evaluation metrics: Precision@$K$, Recall@$K$, and Hit Ratio@$K$. In the experiments, we set $K$ to 10, 20, and 50. For each metric, we first compute the accuracy for each user on the testing data, and then report the averaged accuracy over all users.

\noindent\textbf{Baseline Methods}: We compare CGAT with the following models: (1) \textbf{CFKG}~\cite{ai2018learning} integrates the multi-type user behaviors and item KG into a unified graph, and employs TransE~\cite{bordes2013translating} to learn entity embedding. %The recommendation task is treated as a link prediction task in graph;
(2) \textbf{RippleNet}~\cite{wang2018ripplenet} exploits KG information by propagating a user's preferences over the set of entities along paths in KG rooted at her historical items;
(3) \textbf{MKR}~\cite{wang2019multi} is a multi-task feature learning approach that uses KG embedding task to assist the recommendation task; (4) \textbf{KGNN-LS}~\cite{wang2019knowledge} applies GCN on KG to compute the item embedding by propagating and aggregating the neighborhood information on item KG. %Moreover, the user's personalized preferences on relations and label smoothness regularization are also considered in this model;
(5) \textbf{KGAT}~\cite{wang2019kgat} employs graph attention mechanism on KG to exploit the graph context for recommendation.

\noindent\textbf{Implementation Details}: For CGAT, the dimensionality of latent space $d$ is chosen from $\{8, 16, 32, 64, 128\}$. The number of local neighbors of an entity $S$ and the number of a user's historical items $N$ used in model training are selected from $\{2, 4, 8, 16, 24, 32, 40\}$. The regularization parameters $\lambda_{1}$ and $\lambda_{2}$ are chosen from $\{10^{-6}, 5\times 10^{-6}, 10^{-5}, 5 \times 10^{-5}, 10^{-4}, 5 \times 10^{-4}, 10^{-3}, 10^{-2}\}$. The learning rate $\eta$ is chosen from $\{10^{-4}, 5 \times 10^{-4}, 10^{-3}, 5\times10^{-3}, 10^{-2}\}$. The hyper-parameters of baseline methods are set following original papers. For all methods, optimal hyper-parameters are determined by the performances on the validation data. We implement CGAT by Pytorch, and the Adam optimizer~\cite{kingma2014adam} is used to learn the model parameters.

\begin{table*}
    \caption{Performances of different recommendation algorithms. The best results are in \textbf{bold faces} and the second best results are \underline{underlined}. $^{\ast}$ indicates CGAT significantly outperforms the competitors with $p < 0.05$ using Wilcoxon signed rank significance test.
    }
    %\vspace{-8pt}
    \centering
    \small
    \label{tab:tab1}
    \begin{tabular}{l|l|ccc|ccc|ccc}
        \hline
        Datasets & Methods & P@10 & R@10 & HR@10 & P@20 & R@20 & HR@20 & P@50 & R@50 & HR@50\\\hline
        \multirow{6}{*}{FM}
        % &NFM       & 0.0271          & 0.1120          & 0.2338          & 0.0211          & 0.1792          & 0.3422          & 0.0134          & 0.2777          & 0.4912          \\
        &CFKG      & 0.0280          & 0.1168          & 0.2362          & 0.0222          & 0.1857          & 0.3404          & 0.0135          & 0.2812          & 0.4773        \\
        &RippleNet & 0.0285          & 0.1214          & 0.2423          & 0.0229          & 0.1948          & 0.3628          & 0.0157          & 0.3260          & 0.5336          \\
        &MKR       & 0.0278          & 0.1162          & 0.2356          & 0.0215          & 0.1820          & 0.3356          & 0.0138          & 0.2877          & 0.4809          \\
        &KGNN-LS   & 0.0284          & 0.1186          & 0.2441          & 0.0216          & 0.1824          & 0.3398          & 0.0136          & 0.2828          & 0.4809          \\
        &KGAT      & \underline{0.0466}          & \underline{0.1886}          & \underline{0.3604}          & \underline{0.0341}          & \underline{0.2756}          & \underline{0.4803}          & \underline{0.0206}          & \underline{0.4151}          & \underline{0.6426}          \\
        &CGAT      & \textbf{0.0512}$^{\ast}$ & \textbf{0.2106}$^{\ast}$ & \textbf{0.4022}$^{\ast}$ & \textbf{0.0369}$^{\ast}$ & \textbf{0.2994}$^{\ast}$ & \textbf{0.5203}$^{\ast}$ & \textbf{0.0218}$^{\ast}$ & \textbf{0.4413}$^{\ast}$ & \textbf{0.6687}$^{\ast}$ \\
        \hline
        \multirow{6}{*}{ML}
        % &NFM       & 0.1348          & 0.1359          & 0.6506          & 0.1125          & 0.2144          & 0.7685          & 0.0825          & 0.3700          & 0.8760          \\
        &CFKG      & 0.1054          & 0.1038          & 0.5680          & 0.0896          & 0.1753          & 0.7126          & 0.0633          & 0.2991          & 0.8388          \\
        &RippleNet & 0.1271          & 0.1251          & 0.6227          & 0.1043          & 0.2008          & 0.7474          & 0.0758          & 0.3442          & 0.8667          \\
        &MKR       & 0.1376          & 0.1370          & 0.6581          & 0.1154          & 0.2192          & 0.7765          & 0.0848          & 0.3793          & 0.8852          \\
        &KGNN-LS      & 0.1311          & 0.1310          & 0.6419          & 0.1126          & 0.2172          & 0.7766          & 0.0833          & 0.3762          & 0.8811          \\
        &KGAT      & \underline{0.1533}          & \underline{0.1608}          & \underline{0.7090}          & \underline{0.1274}          & \underline{0.2541}          & \underline{0.8179}          & \underline{0.0910}          & \underline{0.4189}          & \underline{0.9066}          \\
        &CGAT      & \textbf{0.1575}$^{\ast}$ & \textbf{0.1674}$^{\ast}$ & \textbf{0.7219}$^{\ast}$ & \textbf{0.1288}$^{\ast}$ & \textbf{0.2608}$^{\ast}$ & \textbf{0.8264}$^{\ast}$ & \textbf{0.0916}$^{\ast}$ & \textbf{0.4311}$^{\ast}$ & \textbf{0.9191}$^{\ast}$ \\
        \hline
        \multirow{6}{*}{BC}
        &CFKG      & \underline{0.0155}         & 0.0725          & 0.1391          & 0.0101          & 0.0904          & 0.1745          & 0.0061          & 0.1291         & 0.2435          \\
        &RippleNet & 0.0147         & 0.0706          & 0.1336          & 0.0099          & 0.0880           & 0.1736          & 0.0060           & 0.1261         & 0.2429          \\
        &MKR       & 0.0154         & \textbf{0.0732} & 0.1386          & \underline{0.0105}          & \textbf{0.0920}           & \underline{0.1811}          & \underline{0.0063}          & \underline{0.1306}        & \underline{0.2496}          \\
        &KGNN-LS   & \underline{0.0155}         & \underline{0.0730}           & \textbf{0.1411} & 0.0104          & \underline{0.0910}           & 0.1797          & 0.0062          & \underline{0.1306}         & 0.2454          \\
        &KGAT      & 0.0132         & 0.0572          & 0.1202          & 0.0094          & 0.0776          & 0.1600            & \underline{0.0063}          & 0.1172         & 0.2362          \\
        &CGAT      & \textbf{0.0161} & 0.0645  & \underline{0.1402}         & \textbf{0.0119}$^{\ast}$ & \textbf{0.0920} & \textbf{0.1909}$^{\ast}$ & \textbf{0.0078}$^{\ast}$ & \textbf{0.1412}$^{\ast}$ & \textbf{0.2718}$^{\ast}$ \\
        \hline
    \end{tabular}
    %\vspace{-12pt}
\end{table*}

\subsection{Performance Comparison}
Table~\ref{tab:tab1} summarizes the results on different datasets. We make the following observations. On FM and ML datasets, KGAT achieves the best performances among all baselines. On BC dataset, MKR achieves comparable results with KGNN-LS, and outperforms CFKG, RippleNet, and KGAT. The KG and interaction graphs on BC dataset are very sparse. MKR jointly solves the KG embedding and recommendation tasks by learning high-order feature interactions between items and entities. The cross\&compress units are effective to transfer knowledge between the user-item interaction graph and KG, thus can help solve the data sparsity problem.
%The KG and interaction graph on both datasets are very sparse. RippleNet explicitly aggregates multi-hop neighborhood information to represent user preference, which help address the data sparsity problem. Moreover, these results also demonstrate the effectiveness of explicit aggregation of non-local context information in KG.
%MKR achieves better performances than other four baselines on Movielens dataset, which is denser than other two experiment datasets. MKR jointly solves the KG embedding and recommendation tasks, via learning high-order feature interactions between items and entities. The cross\&compress units are more effective to transfer knowledge between the user-item interaction graph and KG, on denser dataset.
Moreover, CGAT usually achieves the best performances on all datasets, in terms of all metrics. In most of the scenarios (\ie 23 among 27 evaluation metrics), the proposed CGAT method significantly outperforms baseline methods with $p<0.05$, using the Wilcoxon signed rank significance test. Over all datasets, on average, CGAT outperforms CFKG, RippleNet, MKR, KGNN-LS, and KGAT by 26.07\%, 21.32\%, 22.29\%, 21.92\%, 9.56\%, respectively, in terms of HR@20. These results demonstrate the effectiveness of CGAT in exploiting both the KG context and users' historical interaction context for recommendation.

\subsection{Ablation Study}

\begin{table}[t]
    \centering
    \small
    %\vspace{-10pt}
    \caption{Performances of CGAT variants estimated by HR@20.}
    %\vspace{-8pt}
    \label{tab:tab4}
    \begin{tabular}{l|c|c|c|c}
    \hline
    Dataset & CGAT\textsubscript{w/o L} & CGAT\textsubscript{w/o G} & CGAT\textsubscript{w/o UA} & CGAT \\ \hline
    FM & 0.5118 & 0.5167 & 0.5136 & \textbf{0.5203} \\
    ML & 0.8193 & 0.8111 & 0.8215 & \textbf{0.8264}\\
    BC & 0.1884 & 0.1864 & 0.1817 & \textbf{0.1909}\\
    \hline
    \end{tabular}
    %\vspace{-12pt}
\end{table}

%To investigate the importance of each component of CGAT,
Moreover, we also conduct ablation studies to evaluate the performances of the following CGAT variants: (1) \textbf{CGAT}\textsubscript{w/o L} deletes the local context embedding of item from CGAT and only considers the non-local context embedding as final context embedding, \ie the coefficient $\sigma(\B{\omega})$ in Eq.\eqref{eq:kgcontextemb} is set to $\B{0}$; (2) \textbf{CGAT}\textsubscript{w/o G} removes the non-local context embedding of item from original model, which is contrast to \textbf{CGAT}\textsubscript{w/o L} model; (3) \textbf{CGAT}\textsubscript{w/o UA} removes the user's embedding in exploiting the local context information in KG (\ie removing $\B{m}_u$ in Eq.~\eqref{eq:attention3}).
%which use item-entity attention mechanism to replace the user-specific attention mechanism, which removes the affect of user embedding in exploration for local graph context.

Due to space limitation, we only report the recommendation accuracy measured by HR@20. We summarize the results in Table~\ref{tab:tab4}, and have the following findings. CGAT consistently outperforms the variants CGAT\textsubscript{w/o L} and CGAT\textsubscript{w/o G}, indicating both local and non-local context in KG are essential for recommendation. CGAT achieves better performance than CGAT\textsubscript{w/o UA}. This demonstrates the user-specific graph attention mechanism is more suitable for personalized recommendation than simple attention mechanism that can not capture users' personalized preferences. CGAT\textsubscript{w/o L} is slightly superior than CGAT\textsubscript{w/o G} on ML and BC datasets. This indicates that non-local context information plays a complementary role to the local context information, and sometimes may be more important than local context information in improving the recommendation accuracy.

\subsection{Parameter Sensitivity Study}
\begin{figure}
\small
\centering
\includegraphics[width=0.95\columnwidth]{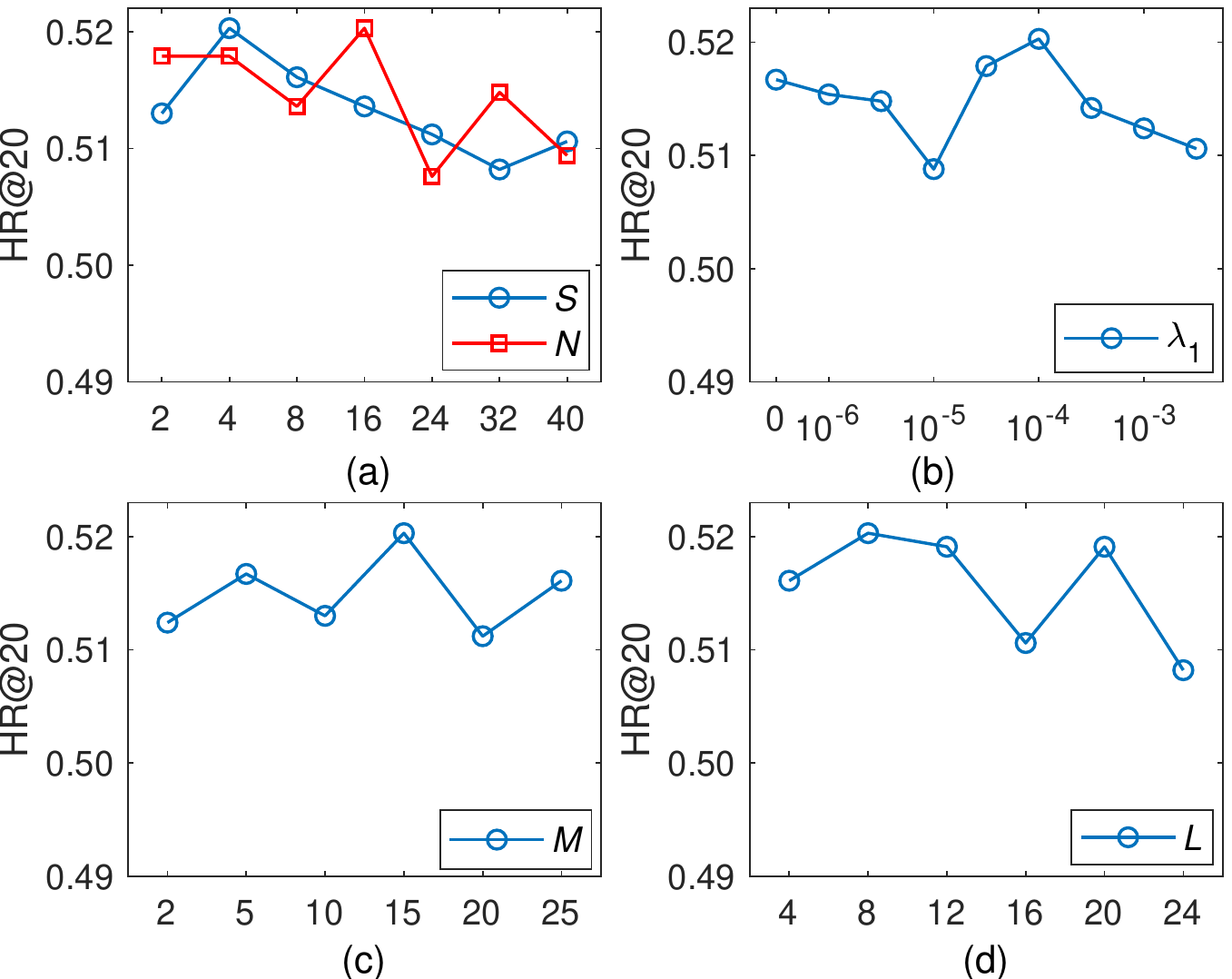}
%\vspace{-8pt}
\caption{Performances of CGAT on FM dataset, w.r.t. different settings of $S$, $N$, $\lambda_1$, $M$, and $L$.}
%\vspace{-13pt}
\label{fig:fig4}
\end{figure}

Figure~\ref{fig:fig4} summarizes the performances of CGAT with respect to (w.r.t.) different settings of key parameters. As the size of neighboring entities in KG usually varies for different items, we study how fixed size of sampled neighbors would affect the performance. From Figure~\ref{fig:fig4}(a), we can note that CGAT achieves the best performance when $S$ is set to 4, while larger $S$ does not help further improve the performance. This optimal setting of $S$ is close to the average number of neighbors of an entity in KG, which is 3.31 on FM dataset. Then, we vary the number of a user's historical items used to represent her potential preferences. As shown in Figure~\ref{fig:fig4}(a), the best performance is achieved by setting $N$ to 16. When $N$ is larger than 16, further increase of $N$ would reduce the performance.
%This optimal setting of $N$ is close to the average number of historical items for each user, which is 11.30 on FM dataset.
Figure~\ref{fig:fig4}(b) shows the performance trend of CGAT w.r.t. different settings of $\lambda_1$. The performances achieved by setting $\lambda_1$ to $5\times10^{-5}$ and $10^{-4}$ are better than that achieved by setting $\lambda_1$ to 0. This observation demonstrates that the KG structure constraint in Eq.~\eqref{eq:kgloss} can help improve the recommendation accuracy. Moreover, we also study the impacts of the number of sampled paths $M$ and the path length $L$ in the BRWS module. From Figure~\ref{fig:fig4}(c), we can note the best performance is achieved by setting $M$ to 15. This indicates the most relevant entities in the non-local neighborhood of an entity can be captured by performing 15 times random walk sampling. As shown in Figure~\ref{fig:fig4}(d), better performance can be achieved by setting $L$ in the range between 4 and 12. Further increasing $L$ causes more training time, however sometimes may cause the decrease in recommendation performances.

%\begin{figure}
%\centering
%\includegraphics[width=0.7\columnwidth]{Dim.pdf}
%\caption{Performances of different methods on Last-FM dataset, with respect to (w.r.t.) different settings of $d$.}
%\label{fig:fig3}
%\end{figure}

\section{Conclusion and Future Work}
This paper proposes a novel recommendation model, called Context-aware Graph Attention Network (CGAT), which explicitly exploits both local and non-local context information in KG and the interaction context information given by users' historical behaviors. Specifically, CGAT aggregates the local context information in KG by a user-specific graph attention mechanism, which captures users' personalized preferences on entities. To incorporate the non-local context in KG, a bias random walk based sampling process is used to extract important entities for the target entity over entire KG, and a GRU module is employed to explicitly aggregate these entity embedding. In addition, CGAT utilizes an item-specific attention mechanism to model the influences between items. The superiority of CGAT has been validated by comparing with state-of-the-art baselines on three datasets. For future work, we intend to develop different aggregation strategies to integrate the context information in KG and interaction graph to improve recommendation accuracy.

\bibliographystyle{named}
\bibliography{ijcai20}

\end{document}